\documentstyle[aps]{revtex}
\newcommand\bm[1]{\mbox{\boldmath$#1$}}
\newcommand{\cth}{\cos  \theta }
\newcommand{\xpA}{\dot{x}_{\tau}^{A}}
\newcommand{\xpB}{\dot{x}_{\tau}^{B}}
\newcommand{\thp}{\dot{\theta}_{\tau}}
\newcommand{\rp}{\dot{r}_{\tau}}
\newcommand{\thpp}{\ddot{\theta}_{\tau}}
\newcommand{\rpp}{\ddot{r}_{\tau}}

\newcommand{\xx}{\dot{x}_{\tau}^{2} }
\newcommand{\yy}{\dot{y}_{\tau}^{2} }
\newcommand{\St}{\Psi^{St}_{\tau}(x)}
\newcommand{\F}{\Psi^{FL}_{\tau}(x)}
\newcommand{\ep}{{\large \epsilon}}
\newcommand{\dw}{\stackrel{o}{{\cal W}}}
\newtheorem{proposition}{Proposition}
\newtheorem{corollary}{Corollary}
\begin{document}
\title{Axially symmetric Einstein-Straus models}
\author{Marc Mars\thanks{Also at Laboratori de F\'{\i}sica Matem\`atica,
Societat Catalana de F\'{\i}sica, IEC, Barcelona.} \\
School of Mathematical Sciences,
Queen Mary and Westfield College, \\
Mile End Road, London E1 4NS, United Kingdom.}
\maketitle
\begin{abstract}
The existence of static and axially symmetric regions in a
Friedman-Lema\^{\i}tre (FL) cosmology is investigated under the only
assumption that the cosmic time and the static time match
properly on the boundary hypersurface. It turns out that the most general
form for the static region is a two-sphere with arbitrarily changing
radius which moves along the axis of symmetry in a determined way.
The geometry of the interior region is completely determined in terms
of background objects. When any of the most widely used energy-momentum
contents for the interior region is imposed, both the interior geometry
and the shape of the static region must become exactly spherically symmetric.
This shows that the Einstein-Straus model, which is the generally accepted
answer for the null influence of the cosmic expansion on the local physics,
is not a robust model and it is rather an exceptional and isolated situation.
Hence, its suitability for solving the interplay between cosmic
expansion and local physics is doubtful and more adequate models should
be investigated.
\end{abstract}
PACS numbers: 04.20.Jb, 04.20.Cv, 98.80.-k, 04.40.Nr
\newpage

\section{Introduction}

The influence of the large scale geometry of the Universe on the local
physics around astrophysical objects is a very old and fundamental
question in gravitational physics. The fact that the geometry
of the Universe as a whole can be approximated very accurately by an
expanding Friedman-Lema\^{\i}tre (FL) model leads naturally to
consider
whether the cosmological expansion has any effect on
the local physics at astrophysical scales.
This type of question was probably first addressed by McVittie \cite{MV}
who used a spherically symmetric spacetime
supposedly describing a point particle immersed in a Friedman-Lema\^{\i}tre
background (see, however, \cite{susm}).
This metric was later analyzed in more detail by
J\"{a}rnefeld \cite{Jar}.
Both authors concluded that the effect of the cosmological expansion
on the planetary orbits is very small and with no measurable
consequences. More recently, Gautreau
\cite{Gau} analyzed the problem by using an inhomogeneous and
spherically symmetric model in
the so-called curvature coordinates \cite{Gau2} (see \cite{Kras} for
a review of these results).
However, the generally accepted answer to this problem
was given by Einstein and Straus in \cite{ES} where  a model
describing a massive point particle
surrounded by a spherically symmetric vacuum region embedded
in a dust FL cosmology was presented
(the now so-called Einstein-Straus model).
The boundary of the vacuum region is a two-sphere
comoving with the cosmological flow.
Thus, Einstein and Straus showed the possibility
of having static regions embedded in an expanding cosmological background
thus implying that the cosmic expansion does not influence the local
physics around the massive object. Despite its clear implications and
simplicity, Einstein and Straus' construction
suffers from several important limitations.
First, the interior geometry of the static region is
exactly spherically symmetric and its boundary is a two-sphere comoving
with the cosmological flow. This a very strong assumption, specially
regarding the  shape of the static cavity.
Second, given the dust background metric,
the radius of the vacuum region surrounding the compact object
is uniquely fixed by its mass. This implies that some situations of
astrophysical interest are not
satisfactorily described by the Einstein-Straus model. Furthermore,
Krasi\'nski has argued \cite{Kras} that the
model is unstable,  meaning that a matching between the vacuum cavity
and the FL background with a slightly bigger radius than the Einstein-Straus
value makes the static region expand with respect to the
cosmological flow (and vice versa for a smaller radius).

Hence, more general models should be analyzed in order to reach a definitive
conclusion on whether  the influence of the cosmological expansion on the
local physics is indeed vanishing and, in that case, on the suitability of the
Einstein and Straus' construction to provide a completely
satisfactory explanation. In order to test
the importance of the spherically symmetric assumption,
other geometries for the interior region
(especially the shape of the boundary) should be considered. Since the local
geometry is generally believed to be static (which is the translation
of the null influence of the background expansion),
the problem can be formulated as to whether non-spherical static cavities
can be embedded in a FL cosmology. In a very recent paper,
Senovilla and Vera \cite{RS} proved the surprising
fact that embedding a cylindrically symmetric static region in an
expanding FL cosmology is {\it always} impossible (irrespective of the
matter contents inside the cavity). This result supports the view that
spherical symmetry may indeed be an essential ingredient for the existence
of the static region.
In order to clarify this question,  we address the existence of
static cavities in a FL background in a rather
general situation. The essential simplifying assumption we make
(otherwise the problem becomes formidable) is that the interior static
region is axially symmetric, both in shape and in spacetime geometry
(no restrictions on the interior energy-momentum contents are imposed).
This assumption is physically very reasonable because
axial symmetry is likely to be present
in many realistic situations. The second assumption we make
is that the two canonical time functions in each region (the
cosmic time in the FL background and the static time in the interior
region) match properly across the
boundary of the static region, so that the global spacetime has a well-defined
and natural time function. This model is general enough to include
both  spherical symmetry and cylindrical symmetry as
particular cases and, hence, its analysis will clarify to what extent
the extreme situations in which the matching is possible
(spherically symmetry) and never feasible (cylindrical
symmetry) are generic.

Considering static regions embedded in an expanding background requires
the use of the theory of matching between spacetimes.
This consists in imposing the
{\it matching conditions} which can be viewed as a set of partial
differential equations to be solved. When the symmetry of the
matching problem is high enough (i.e. when the two spacetimes to be
matched possess at least two isometries which are preserved by the
matching hypersurface), the equations become
ordinary differential equations and the problem simplifies
considerably. Most of the explicit matchings which have been performed
in the literature allow for this type of reduction.
In particular, this is true for the Einstein-Straus model and the
cylindrically symmetric case described above. In our case, however,
the problem is technically  more complex because
the global spacetime possesses {\it only} one isometry,
namely axial symmetry. In order to solve it,
we use a set of {\it constraint
matching conditions}, which are necessary consequences of the matching
conditions  on any spacelike two-dimensional surface
embedded into the matching hypersurface. They appear
naturally when developing
a $2+1$ geometrical decomposition of the matching conditions into
a set of constraints on initial spacelike two-surfaces and a set
of evolution equations determining how the two-surface must evolve
in order to generate the three-dimensional matching hypersurface. The full
decomposition has not been completed yet and will be reported elsewhere.
For the purposes of this paper it will suffice exploiting
the set of constraint matching conditions.

The paper is organised as follows. In section 2, the matching theory
between spacetimes is briefly reviewed and the
constraint matching conditions on spacelike two-dimensional surfaces
are presented.
In section 3, the matching problem between a static and axially
symmetric spacetime and a Friedman-Lema\^{\i}tre expanding cosmology is solved.
Since the problem is technically non-trivial, the procedure is described
in some detail.
First, the constraint matching conditions are imposed. They restrict
the form of the matching hypersurface severely.
It turns out that the most general matching hypersurface consists
of a collection of two-spheres, two-planes or hyperbolic planes centred
on the axis of symmetry and moving arbitrarily along this axis. For
closed FL models only two-spheres are possible; for flat FL models
both two-spheres and two-planes are allowed while for the open case
all three cases are possible. We restrict the analysis to the two-sphere
case because this is the only one
describing a spatially compact region embedded in a FL cosmology.
Then,  the remaining set of matching conditions are imposed. They
determine the interior
geometry of the static region in terms of the background
FL metric and the function $r(t)$ describing how the
radius of the two-spheres generating the matching hypersurface varies
with time. The motion of the set of two-spheres along the axis of
symmetry is also determined by the background metric and $r(t)$. In section 4,
a detailed study of the static and axially symmetric interior
geometry is performed
and a number of important conclusions are obtained. In particular,
the interior metric is shown to be nearly spherically symmetric
with both its
energy-momentum tensor and Petrov type of the same type as in the
exact spherically symmetric case. Furthermore, analyzing the most
widely used energy-momentum tensors for the interior
region (vacuum, $\Lambda$-term, perfect fluid and electrovacuum) we show
that the interior geometry must be {\it exactly} spherically symmetric in all
those cases and the boundary of the cavity must be generated by a two-sphere
with its centre being at rest with respect to the cosmological flow.
Hence, we conclude that
the Einstein-Straus model is essentially unique (when axial
symmetry is imposed) and that the spherical symmetry assumption is
indeed crucial for the existence of static cavities inside a FL cosmology.
Consequently, obtaining a robust model determining whether or not
the cosmological
expansion has any influence on the local physics
is still an open
problem which deserves further analysis and investigation.

\section{Junction of spacetimes and constraint matching conditions.}

The matching (see \cite{MS} and references therein)
between two spacetimes with boundary,
$\left(V^{+}, g^{+} \right)$ and $\left (V^{-},g^{-}
\right)$ with corresponding boundaries
$\Omega^{+}$ and $\Omega^{-}$, consists in constructing a spacetime
manifold $\left (V,g \right)$ by identifying the points,
(and the tangent spaces) on the two boundaries,
which are therefore called {\it matching hypersurfaces}.
The point-to-point identification requires the
existence of a diffeomorphism between
$\Omega^{+}$ and $\Omega^{-}$. Therefore, they can be also
considered as diffeomorphic to an abstract three-dimensional
manifold $\Omega$ so that there exist two embeddings
\[
\Phi^{+} : \, \, \, \Omega \,  \longrightarrow  \, \, \,V^{+} ,
\hspace{2cm} \Phi^{-} : \, \, \, \Omega \,  \longrightarrow  \, \, \,V^{-},
\]
satisfying $\Phi^{+} \left (\Omega \right) = \Omega^{+}$ and
$\Phi^{-} \left (\Omega \right) = \Omega^{-}$.
Using local coordinates
$\left \{\xi^{i} \right \}$ for $\Omega$ ($i,j,\cdots=1,2,3$),
$\left \{x_{+}^{\alpha} \right \}$
for $V^{+}$ ($\alpha, \beta, \cdots = 0,1,2,3$) and $\left \{x_{-}^{\alpha} \right \}$
for $V^{-}$ the two embeddings take a local form
\[
\Phi^{+} : \, \, \,
\xi^{i} \longrightarrow x_{+}^{\alpha}=x_{+}^{\alpha} \left (\xi \right ),
\hspace{2cm}
\Phi^{-} : \, \, \,
\xi^{i} \longrightarrow x_{-}^{\alpha}=x_{-}^{\alpha} \left (\xi \right ),
\]
where $x_{+}^{\alpha} \left (\xi \right )$ and
$x_{-}^{\alpha} \left (\xi \right )$ are functions of $\xi^i$.
As an important theorem by Clarke and Dray \cite{CD} shows,
the construction of a spacetime $V = V^{-}\cup V^{+}$
with continuous metric $g$ can be performed if and only if
the two first fundamental forms
${\Phi^{+}}^{*} \left (g^{+} \right )$ and
${\Phi^{-}}^{*} \left (g^{-} \right)$
on $\Omega$ coincide ($\Phi^{*}$ denotes the pull-back of the map $\Phi$). In local coordinates, these conditions read
\begin{equation}
\left . g^{+}_{\alpha\beta} \right |_{
x_{+} \left (\xi \right)}
\frac{\partial x_{+}^{\alpha}}{\partial \xi^{i}}
\frac{\partial x_{+}^{\beta}}{\partial \xi^{j}}
=\left . g^{-}_{\alpha\beta} \right |_{
x_{-} \left (\xi \right)}
\frac{\partial x_{-}^{\alpha}}{\partial \xi^{i}}
\frac{\partial x_{-}^{\beta}}{\partial \xi^{j}}, \label{Mat1}
\end{equation}
and will be called {\it preliminary matching conditions}.
When (\ref{Mat1}) is fulfilled,
the Riemann tensor in $\left(V,g\right )$
is well-defined in a distributional sense. In general, this
distribution has a Dirac delta part  with support on the matching
hypersurface, which can be used, when
appropriate, to describe thin
shells of matter and/or pure gravitational field. Whenever these physical
objects are absent, the Dirac delta contribution must vanish and the
{\it second set of matching conditions} must be imposed. First,
two transverse vector fields, $\vec{l}_{+}$
on $\Omega^{+}$ and $\vec{l}_{-}$ on $\Omega^{-}$, (called {\it rigging}
vector fields) must be chosen. As they prescribe how the tangent spaces are to
be identified, their choice must be compatible with the
continuity of the metric in $V$. This demands
\begin{eqnarray}
\left . g^{+}_{\alpha\beta} \, l_{+}^{\alpha} \,
 l_{+}^{\beta} \right |_{x_{+}\left (\xi\right)} =
\left . g^{-}_{\alpha\beta}  \, l_{-}^{\alpha} \,
l_{-}^{\beta} \right |_{x_{-} \left ( \xi \right) }, \nonumber \\
\left . g^{+}_{\alpha\beta} \, l^{\alpha}_{+} \right |_{x_{+} \left (\xi\right)
 }
\frac{\partial x_{+}^{\beta}}{\partial \xi^{j}}  =
\left . g^{-}_{\alpha\beta} \,  l^{\alpha}_{-} \right |_{x_{-} \left (\xi
\right)}
 \frac{\partial x_{-}^{\beta}}{\partial \xi^{j}}. \label{Conrig}
\end{eqnarray}
The two riggings $\vec{l}_{+}$ and $\vec{l}_{-}$ must
have different relative orientations with respect
to $\Omega^{+}$ and $\Omega^{-}$, respectively (i.e.
either $\vec{l}_{+}$ points outside $V^{+}$ everywhere
and $\vec{l}_{-}$ points inside $V^{-}$ everywhere or vice versa).
Once the riggings are chosen, the matching conditions demand that the pull-back
of the covariant derivative of the rigging one-form (obtained by
lowering the index to $\vec{l}\,$) coincides when calculated using
any of both embeddings. In local coordinates,
\begin{eqnarray}
\left . \frac{}{} l_{+\beta} \right |_{x_{+} \left (\xi \right )}
 \frac{\partial x_{+}^{\alpha}}{\partial \xi^{i}} \nabla^{+}_{
\alpha} \left ( \frac{\partial x_{+}^{\beta}}{\partial \xi^{j}} \right) =
\left . \frac{}{}l_{-\beta} \right |_{x_{-} \left (\xi \right) }
 \frac{\partial x_{-}^{\alpha}}{\partial \xi^{i}} \nabla^{-}_{
\alpha} \left ( \frac{\partial x_{-}^{\beta}}{\partial \xi^{j}} \right)
\label{Mat2} \hspace{1cm} \Longleftrightarrow \\
\left . \frac{}{} l_{+\beta} \right |_{x_{+} \left (\xi \right )}
\left (
\frac{\partial^2 x_{+}^{\beta}}{\partial \xi^i \partial \xi^{j}}
+ \Gamma^{+ \beta}_{\mu \nu} \frac{\partial x_{+}^{\mu}}{\partial \xi^{i}}
\frac{\partial x_{+}^{\nu}}{\partial \xi^{j}} \right ) =
\left . \frac{}{} l_{-\beta} \right |_{x_{-} \left (\xi \right )}
\left (
\frac{\partial^2 x_{-}^{\beta}}{\partial \xi^i \partial \xi^{j}}
+ \Gamma^{- \beta}_{\mu \nu} \frac{\partial x_{-}^{\mu}}{\partial \xi^{i}}
\frac{\partial x_{-}^{\nu}}{\partial \xi^{j}} \right ), \nonumber
\end{eqnarray}
where $\nabla^{+}$ ($\nabla^{-}$) denotes the covariant derivative
and $\Gamma^{+ \beta}_{\mu \nu}$ ($\Gamma^{-\beta}_{\mu \nu}$)
are the corresponding Christoffel symbols in  $V^{+}$ ($V^{-}$).

In many physically interesting
problems the degree of symmetry is high enough so that
the conditions (\ref{Mat1}) and (\ref{Mat2})
become a system of ordinary differential equations, which
simplifies the problem considerably. In general,
however, this reduction does not happen
and the partial differential system must be considered. As we shall see
below, some of these cases can be handled more easily by using a $2+1$
decomposition of the matching conditions. This consists in
obtaining a set of geometrical conditions on an initial
two-dimensional surface and evolution equations
determining how this surface must evolve in order to
generate the matching hypersurface. This theory will be
reported elsewhere when fully developed. For the aim of this paper,
only the {\it constraint matching conditions} will be necessary.
In order to describe them, some notation must be introduced.

The main assumption for the $2+1$ decomposition
is that the hypersurface $\Omega$ can be foliated by a set
of two-dimensional surfaces, which will be denoted by $\left \{
S_{\tau} \right \}$ ($\tau$ is a parameter identifying the element
of the foliation and plays the role of an evolution parameter). Each
two-surface in the foliation can be immersed into $\Omega$ by using the
natural inclusion
\[
i_{\tau} : \, \, \, S_{\tau} \, \longrightarrow \, \, \, \Omega.
\]
The surface $S_{\tau}$ can then be embedded into
$V^{+}$ and $V^{-}$, respectively, by the maps
\begin{equation}
\Phi^{+}_{\tau} \equiv \Phi^{+} \circ i_{\tau}, \hspace{2cm}
\Phi^{-}_{\tau} \equiv \Phi^{-} \circ i_{\tau} \label{defPhi}
\end{equation}
which define the two image surfaces $S^{+}_{\tau} \equiv
\Phi^{+}_{\tau} \left (S_{\tau} \right)$ and $S^{-}_{\tau} \equiv
\Phi^{-}_{\tau} \left (S_{\tau} \right)$. We will further
assume that both $S^{+}_{\tau}$ and $S^{-}_{\tau}$ are {\it spacelike}
everywhere.
Let us consider a point $x \in S_\tau$ and define the
vector space
\[
\left (T_{x} S_{\tau}^{+} \right )_{\bot} \equiv \left \{ \bm{n}
\in T^{\star}_{\Phi^{+}_{\tau} \left(x \right)} V^{+} \, \, ; \, \,
\bm{n} \left (d\Phi^{+}_{\tau} \left (T_x S_{\tau} \right) \right) = 0
\right \},
\]
where
$d\Phi_{\tau}^{+}$ is the differential application of the embedding
$\Phi_{\tau}^{+}$, and $T_{y} M$, $T_{y}^{\star} M$ denote,
respectively, the tangent and cotangent
spaces of a manifold $M$ at any point $y \in M$.
The vector space  $\left (T_{x} S_{\tau}^{+} \right )_{\bot}$ is simply
the set of normal one-forms to the two-surface $S^{+}_{\tau}$
in $V^{+}$.
Since $S^{+}_{\tau}$ is assumed to be spacelike,
$\left (T_{x} S_{\tau}^{+} \right )_{\bot}$ is
two-dimensional and timelike. The second fundamental form of
$S^{+}_{\tau}$ along the direction $\bm{n}$ is defined, as usual, as
\[
\bm{K^{+}_{S_{\tau}}} \left (\bm n \right )
\equiv {\Phi^{+}_{\tau}}^{\star} \left ( \nabla^{+}
\bm{n} \right).
\]
where $\bm{n}$ is a one-form field orthogonal to $S^{+}_{\tau}$ everywhere
(the set of all these objects will be denoted by
$\left (T S_{\tau}^{+} \right )_{\bot}$). Obviously, similar definitions hold
by interchanging plus and minus everywhere.

The {\it constraint matching conditions} on $S_{\tau}$ are essentially
the restrictions of the full set of matching conditions into the two-surface
$S_{\tau}$. They can also be splitted into two different sets,
one involving first fundamental forms and another involving second fundamental
forms. The first one is an immediate consequence of (\ref{Mat1}) and the
definition (\ref{defPhi}) and demands, quite naturally,
the equality of the two first fundamental forms on $S_{\tau}$.
Explicitly,
\begin{equation}
{\Phi^{+}_{\tau}}^{\star}  \left (g^{+} \right ) =
{\Phi^{-}_{\tau}}^{\star} \left (g^{-} \right ). \label{ConstMat1}
\end{equation}
The second set of constraint matching conditions is somewhat more
involved, although its geometrical contents can be clearly understood.
The matching of the two spacetimes, requires, as mentioned above, the
identification of the tangent spaces at the points on the
matching hypersurfaces. Furthermore, this identification must
be done so that the resulting metric tensor is continuous.
When working with the foliation $\{ S_{\tau} \}$, part of the job is
accomplished by identifying the tangent planes to the
two surfaces $S^{+}_{\tau}$ and $S^{-}_{\tau}$. The remaining
identification requires the existence of a linear and isometric application
\begin{equation}
\xi_{\tau}^{x} :\, \, \, \left (T_{x} S_{\tau}^{+} \right )_{\bot}
 \,  \hspace{3mm} \longrightarrow \hspace{3mm}  \, \, \,
\left (T_{x} S_{\tau}^{-} \right )_{\bot}, \label{defxi}
\end{equation}
which completes the identification of the tangent spaces at the
points $\Phi^{+}_{\tau}(x)$ and $\Phi^{-}_{\tau}(x)$.
It is not difficult to show that the preliminary and
second set of matching conditions can be combined to imply the following
condition on $S_{\tau}$
\begin{equation}
\bm{K^{+}_{S_{\tau}}} \left (\bm{n} \right ) =
\bm{K^{-}_{S_{\tau}}} \left (\xi_{\tau} \left ( \bm{n} \right ) \right ),
\hspace{2cm} \forall \,
\bm{n} \, \in \, \left ( T S^{+}_{\tau} \right )_{\bot},
\label{ConstMat2}
\end{equation}
where the  field $\xi_{\tau} \left ( \bm{n} \right )$
is defined, naturally,
as  $\left . \xi_{\tau} \left ( \bm{n} \right ) \right
|_{\Phi^{-}_{\tau}(x)} = \xi_{\tau}^{x}
\left ( \left . \bm{ n} \right |_{\Phi^{+}_{\tau}(x) } \right )$,
$\forall x \in S_{\tau}$. The existence of an isometric application
(\ref{defxi}) satisfying (\ref{ConstMat2}) constitutes the
{\it second set of constraint matching conditions}.
Due to the linearity of $\xi_{\tau}^{x}$, and
the ${\cal F}$-linearity of $\bm{K^{-}_{S_{\tau}}}$ and
 $\bm{K^{+}_{S_{\tau}}}$ (i.e. $\bm{K^{\pm}_{S_{\tau}}} (f \bm{n} ) =
f \bm{K^{\pm}_{S_{\tau}}} ( \bm{n} )$ for any scalar function $f$),
it suffices
to impose (\ref{ConstMat2}) for two
one-form fields  $\bm{n_1}, \bm{n_2} \in
\left ( T S^{+}_{\tau} \right )_{\bot}$ being
linearly independent at each
point on $S^{+}_{\tau}$. Let us now write down the local form of the
constraint matching conditions (\ref{ConstMat1}) and (\ref{ConstMat2}).
Introducing local coordinates $\{\zeta^{A} \}$
($A,B, \cdots=1,2$) in the two-surface $S_{\tau}$,  the two embeddings read,
locally,
\[
\Phi^{+}_{\tau} : \, \, \,
\zeta^{A} \longrightarrow x_{+}^{\alpha}=x^{+\alpha}_{\tau}
\left (\zeta \right ),
\hspace{2cm}
\Phi^{-}_{\tau} : \, \, \,
\zeta^{A} \longrightarrow x_{-}^{\alpha}= x^{-\alpha}_{\tau}
 \left (\zeta\right ),
\]
where $x^{+\alpha}_{\tau} \left (\zeta \right )$ and
$x^{-\alpha}_{\tau} \left (\zeta \right )$ are functions of
$\zeta^A$. The condition (\ref{ConstMat1}) becomes, obviously,
\begin{equation}
\left . g^{+}_{\alpha\beta} \right |_{
x^{+}_{\tau} \left (\zeta \right)}
\frac{\partial x^{+\alpha}_{\tau}}{\partial \zeta^{A}}
\frac{\partial x^{+\alpha}_{\tau}}{\partial \zeta^{B}}
=\left . g^{-}_{\alpha\beta} \right |_{
x^{-}_{\tau} \left (\zeta\right)}
\frac{\partial x^{-\alpha}_{\tau}}{\partial \zeta^{A}}
\frac{\partial x^{-\alpha}_{\tau}}{\partial \zeta^{B}}. \label{CMat1}
\end{equation}
Regarding (\ref{ConstMat2}), we take arbitrary bases,
$\left . \bm{n^{A}_{+}} \right |_{\Phi^{+}_{\tau}(x)}$ of $
\left (T_{x} S_{\tau}^{+} \right )_{\bot}$ and
$\left . \bm{n^{A}_{-}} \right |_{\Phi^{-}_{\tau}(x)}$ of $
\left (T_{x} S_{\tau}^{-} \right )_{\bot}$, and define
the scalar quantities
\[
\gamma_{+}^{AB}(x) \equiv g^{+} \left .
\left (\bm{n^{A}_{+}},
\bm{n^{B}_{+}} \right) \right |_{\Phi^{+}_{\tau}(x)}, \hspace{1cm}
\gamma_{-}^{AB}(x) \equiv g^{-} \left .
\left (\bm{n^{A}_{-}},
\bm{n^{B}_{-}} \right) \right |_{\Phi^{-}_{\tau}(x)}.
\]
The linear isometric application $\xi^{x}_{\tau}$ takes the form
\[
\xi^{x}_{\tau} \left ( \left .
\bm{n^{A}_{+}} \right |_{\Phi^{+}_{\tau}(x)} \right) =
{\xi_{\tau}}^{A}_{B}(x) \left . \bm{n^{B}_{-}} \right |_{\Phi^{-}_{\tau}
(x)},
\]
where ${\xi_{\tau}}^{A}_{B}(x)$ are a set of scalar functions defined on
$S_\tau$ which must satisfy
\[
{\xi_{\tau}}^{A}_{B}(x) {\xi_{\tau}}^{C}_{D}(x)\gamma_{-}^{BD}(x) =
\gamma_{+}^{AC}(x)
\]
to ensure that $\xi_{\tau}^{x}$ is an isometry.
The condition
(\ref{ConstMat2}) can now be rewritten as
\begin{equation}
\left . \frac{}{} n^{C}_{+\beta}
\left (
\frac{\partial^2 x^{+\beta}_{\tau}}{\partial \zeta^A \partial \zeta^{B}}
+ \Gamma^{+ \beta}_{\mu \nu} \frac{\partial x^{+\mu}_{\tau}}{\partial
\zeta^{A}}
\frac{\partial x^{+\nu}_{\tau}}{\partial \zeta^{B}} \right )
 \right |_{x^{+}_{\tau} \left (\zeta \right )} =
\left . \frac{}{} {\xi_{\tau}}^{C}_{D}(\zeta) n^{D}_{-\beta}
\left (
\frac{\partial^2 x^{-\beta}_{\tau}}{\partial \zeta^A \partial \zeta^{B}}
+ \Gamma^{- \beta}_{\mu \nu} \frac{\partial x^{-\mu}_{\tau}
}{\partial \zeta^{A}}
\frac{\partial x^{-\nu}_{\tau}}{\partial \zeta^{B}} \right )
\right |_{x^{-}_{\tau} \left (\zeta \right )}.
\label{CMat2}
\end{equation}
In the next section, the conditions
(\ref{CMat1}) and (\ref{CMat2}) will be exploited to
restrict the form of the
matching hypersurface in the axially symmetric Einstein-Straus model.
They will allow for a complete resolution of the matching problem we are
considering.

\section{Axially symmetric Einstein-Straus model.}

Let us now study the main object of this paper, namely the
existence of static, axially symmetric regions embedded
in a Friedman-Lema\^{\i}tre universe.
We choose standard spherical coordinates in the
FL cosmology so that the line-element takes the usual form
\begin{equation}
ds^2_{FL} = -dt^2 + a^2(t) \left [ dr^2 + \Sigma^2\left (r,\epsilon \right)
\left ( d\theta^2 +   \sin^2 \theta d \phi^2 \right )
\right ], \label{FLmetric}
\end{equation}
where $a(t)$ is the scale factor of the cosmological model (we do not
restrict its form whatsoever a priori) and $\Sigma \left (r,\epsilon \right)$
is the standard function
\begin{equation}
\Sigma \left (r, \epsilon \right) = \left \{
\begin{array}{cl}
\sinh r &  \mbox{if } \, \, \epsilon = -1 \\
r & \mbox{if } \, \, \epsilon =0 \\
\sin r & \mbox{if } \, \, \epsilon =1
\end{array}
\right . .
\end{equation}
Regarding the static, axially symmetric spacetime, we will make the
usual assumption that it is orthogonally
transitive (i.e. the two-planes orthogonal
to the isometry group orbits are surface-forming). As an important
theorem by Kundt and Tr\"{u}mper
\cite{Ktru} shows
(generalizing Papapetrou's theorem for vacuum \cite{Papa}),
this condition is automatically satisfied for a wide
class of energy-momentum tensors including vacuum, $\Lambda$-term (i.e.
cosmological constant), perfect fluids and electrovacuum. Thus, the
orthogonally transitive assumption does not represent a severe restriction
on the  kind of regions we are considering. Under this assumption, there
exist coordinates in which the line-element in the static
region reads
\begin{equation}
ds^2_{St}= - F^2 \left (x^A \right ) dT^2
+ g_{AB} \left (x^{C} \right) dx^A dx^B +
W^2 \left (x^A \right) d \hat{\phi}^2, \label{MetEA}
\end{equation}
so that $\partial_T$ is the static and $\partial_{
\hat{\phi}}$
is the axial Killing vector. The coordinate transformations keeping the
structure of (\ref{MetEA}) unchanged are
\begin{equation}
T \rightarrow  \alpha_{1} T + \alpha_0, \hspace{1cm}
x^{A} \rightarrow x^{A} \left (\tilde{x}^A \right),
\hspace{1cm} \hat{\phi} \rightarrow \hat{\phi}+c,
\label{freedSt}
\end{equation}
where $\alpha_1 \neq 0$, $\alpha_0$ and $c$
are arbitrary constants and $x^{A} \left (\tilde{x}^A \right)$ are
arbitrary functions with non-vanishing Jacobian.

Both FL and static spacetimes admit canonical foliations
by spacelike hypersurfaces.
In the FL cosmology, this foliation
is defined by the homogeneity and isotropy hypersurfaces
$t=\mbox{const}$. In the
static region,  the foliation is given by the hypersurfaces
orthogonal to the static Killing vector, $T= \mbox{const}$.
Since the physical
problem we treat is the existence of static regions living
in a cosmological background, we can assume that the matching hypersurfaces
are not  tangent to any of the hypersurfaces $t=\mbox{const.}$ in the
FL spacetime or $T=\mbox{const.}$ in the static spacetime. This
case would rather correspond to a phase transition between a static
spacetime and an expanding FL cosmology, which is a
completely different situation.
Consequently, the foliation defined by the cosmic
time induces a canonical foliation in the matching hypersurface
$\Omega^{FL}$,
defined by $S^{FL}_{\tau} = \Omega^{FL} \cap \left \{
t= \tau \right \}$. Similarly, the matching hypersurface in the static
spacetime $\Omega^{St}$
(the labels $FL$ and $St$ will replace
the scripts plus and minus used in the previous section) can be
canonically foliated by the two-surfaces $S^{St}_{{\cal T}} =
\Omega^{St} \cap \left \{T =  {\cal T}\right \}$. Hence, we have that each
region in the glued spacetime has a natural and canonical time
function, the cosmic time in the FL background and the static
time in the interior region. It is natural from the physical point of
view to demand that these two canonical times coincide on the
matching hypersurface. Here, we make
the second fundamental assumption in this paper, namely, we demand
that the global Einstein-Straus spacetime possesses also
a canonical time function which coincides with the cosmic time in
the FL background and with the static time in the interior
region\footnote{If the
interior region possesses a high enough degree of symmetry (e.g. Minkowski
spacetime) the static time may be not unique. However, since we did not
specify the form of the interior metric (\ref{MetEA}) a priori,
we are not choosing between any of them (whenever there is more than one)
and therefore they are all treated on the same footing.
In this particular case, the assumption we are making
is that there exists {\it at least one} static
time which matches appropriately with the cosmic time on the matching
hypersurface. As we shall see below, the matching problem will fix
the interior geometry completely and, therefore, it will choose which
of the static times (if there is more than one) is the appropriate one
to match with the exterior FL background.}.
If this assumption were violated, then
spacetime events on the boundary of the static region being
equally old with respect to the cosmic time, would correspond to different
instants of time with respect to the static time, which is an
undesirable property for the model we are constructing. Notice that
this condition we are imposing
{\it is} satisfied in both the spherically
symmetric Einstein-Straus model and in the static cylindrically
symmetric model in \cite{RS}. Therefore, the assumption is not only physically
plausible but it is also general enough so that a bridge
between the two extreme cases is constructed. Hence, analyzing this
case will certainly determine  whether the spherically symmetric
Einstein-Straus model is exceptional or not.
It must be emphasized, however, that we are indeed making a simplifying
assumption which is not mathematically necessary a priori. It would be
interesting analyzing the problem in full generality, although this
is probably a much more difficult question.

The condition we are imposing can be translated in mathematical terms by
demanding that the two canonical foliations in the matching hypersurface
coincide in the glued manifold. In other words, we assume there
exists a foliation $\left \{ S_{\tau} \right \}$
in the abstract matching hypersurface $\Omega$ such that the two
embeddings defining the matching hypersurfaces
\[
\Phi^{FL} : \, \, \, \Omega \,  \longrightarrow  \, \, \,V^{FL} ,
\hspace{2cm} \Phi^{St} : \, \, \, \Omega \,  \longrightarrow  \, \, \,
V^{St},
\]
satisfy $\Phi^{FL} \left (S_\tau \right) = S^{FL}_{\tau}$ and
$\Phi^{St} \left (S_{\tau} \right ) = S^{St}_{T(\tau)}$ (the geometrical
coincidence of the foliation allows
for a reparametrisation $T(\tau)$).

Regarding the condition that the static region is axially symmetric, this
implies not only that the interior metric possesses an axial Killing vector,
but also that the two matching
hypersurfaces preserve this symmetry. More precisely, they must be
invariant under the action of the axial symmetry  in each spacetime.
Since the two spacelike foliations in $V^{FL}$ and $V^{St}$ are
also axially symmetric, it follows that
each two-surface $S_{\tau}$ admits an axial Killing vector field. Choosing
coordinates $\{ \lambda, \varphi \}$ on $S_{\tau}$ adapted to this
axial symmetry (i.e. such that the axial Killing reads $\partial_{\varphi}$),
the most general
form for the two embeddings of $S_{\tau}$ into $V^{FL}$ and $V^{St}$
can be easily seen to be
\begin{eqnarray}
\Phi^{FL}_{\tau} : \, \, \, S_{\tau} \,&  \longrightarrow & \, \, \,V^{FL}
\nonumber \\
\Phi^{FL}_{\tau} : \, \left \{ \lambda, \varphi \right \}
\, & \longrightarrow & \, \left \{
         t= \tau, r = r_{\tau} \left ( \lambda \right),
         \theta = \theta_{\tau} \left ( \lambda \right ), \phi = \varphi
\right \},  \label{imbFL} \\
\nonumber \\
\Phi^{St}_{\tau} : \, \, \, S_{\tau} \, & \longrightarrow & \, \, \,V^{St}
\nonumber \\
\Phi^{St}_{\tau} : \, \left \{ \lambda, \varphi \right \}
\, & \longrightarrow & \, \left \{
            T=  T(\tau),  x^{A} = x^{A}_{\tau} \left (\lambda \right ),
            \hat{\phi} = \varphi \right \}, \label{imbES}
\end{eqnarray}
where $r_{\tau} \left ( \lambda \right )$, $\theta_{\tau} \left ( \lambda
 \right )$, $x^{A}_{\tau} \left (\lambda \right )$and $ T(\tau)$
are unknown functions. Imposing the first constraint matching
conditions (\ref{ConstMat1}) gives the two relations
\[
\left . \frac{}{}W(x^{A}) \right |_{\St} =
\left . \frac{}{} a \Sigma \sin \theta \right |_{\F}, \hspace{4mm}
\xx \equiv
\left . \frac{}{} g_{AB}  \xpA \xpB \right |_{\St}= \left . \frac{}{}
a^2 \left (  \rp^2 +  \Sigma^2  \thp^2 \right) \right |_{\F},
\]
where the dot means derivative with respect to $\lambda$. Regarding
the second set of conditions (\ref{ConstMat2}), we must first
identify the  normal planes to the two-surface $S_{\tau}$ in each
spacetime and then construct the most general linear isometry
between them. The vector space $\left (T_{x} S_{\tau}^{St} \right )_{\bot}$
is spanned by the two mutually orthogonal one-forms
$$
\left . \frac{}{}  dT \right |_{\Phi^{St}_{\tau}(x)},
\hspace{1cm} \left . \bm{\kappa}  \frac{}{}\right |_{\St}
\equiv \ep_{AB} \xpA \left . dx^{B} \right |_{\Phi^{St}_{\tau}(x)},
$$
where $\ep_{AB}$ is the totally antisymmetric symbol with
$\epsilon_{12} = 1$.
Similarly, the vector space $\left (T_{x} S_{\tau}^{FL} \right )_{\bot}$
is spanned by the two mutually orthogonal one-forms
\[
\left . \frac{}{} dt \right |_{\Phi^{FL}_{\tau}(x)}, \hspace{1cm}
 \left . \frac{}{}\bm{\alpha} \right |_{\Phi^{FL}_{\tau}(x)}
 \equiv  \left . \frac{}{} \thp  dr  - \rp  d\theta \right |_{\Phi^{FL}_{\tau}(x)}.
\]
Thus, the most general linear and isometric application $\xi^{x}_{\tau} : \, \,
\left (T_x
S^{FL}_{\tau} \right )_{\bot} \, \, \longrightarrow \, \,
\left (T_x  S^{St}_{\tau} \right )_{\bot}$ is given by the linear extension of
\begin{eqnarray*}
\xi^{x}_{\tau}  \left (\left . dt \right |_{\F} \right )=
\left . \frac{}{}
\eta_{1} F \cosh \beta  dT  + \eta_{2} \sqrt{\frac{G}{\xx}}
\sinh \beta  \, \, \bm{\kappa}
 \right |_{\St}, \\
\xi^{x}_{\tau}
\left (\left . \frac{a^2 \Sigma}{\sqrt{\yy}} \bm{\alpha}
 \right |_{\F} \right )= \left . \frac{}{}
\eta_{1} F \sinh \beta  dT  + \eta_{2} \sqrt{\frac{G}{\xx}} \cosh \beta
\, \,  \bm{\kappa}  \right |_{\St},
\end{eqnarray*}
where $\beta \left (\lambda,\phi \right)$
is an arbitrary function,
$\eta_{1}$ and $\eta_{2}$ are arbitrary signs and we have defined
$\yy \equiv \left .
a^2 \left ( \rp^{2} + \Sigma^2 \thp^{2} \right )
\right |_{\Phi^{FL}_{\tau}(x)}$ and $G \equiv \det \left(
g_{AB} \right)$. A consequence of the staticity of the interior region
is that $\bm{K^{St}_{S_{\tau}}} \left (\left . dT
\right |_{\Phi^{FL}_{\tau}(x)} \right) =0$. Consequently, the
second set of matching conditions (\ref{ConstMat2}) becomes
\begin{eqnarray*}
\left. \begin{array}{rr}
{\displaystyle
\bm{K^{FL}_{S_{\tau}}} \left (\left . dt \right |_{\F} \right ) =
\eta_{2} \sqrt{\frac{G}{\xx}} \sinh \beta \bm{K^{St}_{S_{\tau}}}
\left ( \left .  \bm{\kappa} \right |_{\St} \right) }\\
\\
{\displaystyle
 \frac{a^2 \Sigma}{\sqrt{\yy}} \bm{K^{FL}_{S_{\tau}}} \left (\left .
\bm{\alpha} \right |_{\F} \right ) =
\eta_{2} \sqrt{\frac{G}{\xx}} \cosh \beta \bm{K^{St}_{S_{\tau}}}
\left ( \left .  \bm{\kappa} \right |_{\St} \right)}
\end{array} \right \}
\end{eqnarray*}
which implies
\begin{equation}
\bm{K^{FL}_{S_{\tau}}} \left (\left . dt \right |_{\F} \right ) =
\frac{a^2 \Sigma}{\sqrt{\yy}} \tanh \beta
\bm{K^{FL}_{S_{\tau}}} \left (\left .
\bm{\alpha} \right |_{\F} \right ) . \label{FLobj}
\end{equation}
This equation involves only objects from the FL background metric. A trivial
calculation shows that $\bm{K^{FL}_{S_{\tau}}} \left (\left . dt
\right |_{\F} \right )$ and $\bm{K^{FL}_{S_{\tau}}} \left (\left .
\bm{\alpha} \right |_{\F} \right )$
are both diagonal and therefore (\ref{FLobj})
implies two conditions. Combining them
so that the hyperbolic angle $\beta$ disappears, we obtain the equation
\begin{equation}
\left .
-\thp \rpp + \rp \thpp + \frac{\Sigma_{,r}}{\Sigma}
{\rp}^2 \thp + \left (
\frac{\rp^2}{\Sigma^2}  + \thp^2 \right ) \frac{\cos \theta}{\sin \theta}
\rp   \right |_{S^{FL}_{\tau}} =0, \label{DefEq}
\end{equation}
where we have dropped  a global factor $a_{,t}$ (comma means, as usual,
partial derivative) which is non-zero
because the cosmological model is assumed to be expanding (our
aim is precisely analyzing the possible effect of the cosmological
expansion in the local physics).
(\ref{DefEq}) is an ordinary differential equation which
defines
the possible shapes of $S_{\tau}^{FL}$. It can be explicitly solved to
give
\begin{equation}
\left .
\frac{}{} \Sigma \left ( r ,\epsilon \right) \right |_{r= r_{\tau}
\left (\lambda \right)}  = \left . \frac{2\sigma_1
\left (\sigma_0 \cos \theta + \sqrt{1-\sigma_0^2 \sin^2 \theta}
\right)}{\left ( \sigma_0 \cos \theta + \sqrt{1-\sigma_0^2 \sin^2 \theta}
\right)^2 + \epsilon \sigma_1^2}  \right |_{\theta = \theta_{\tau} \left (
\lambda \right)},
\label{stau}
\end{equation}
where $\sigma_1$ and $\sigma_0$ are arbitrary integration constants
(the particular solution $\theta= \pi/2$ can be obtained from this
expression by putting $\sigma_0 = 1$ and $\sigma_1=0$ and following
an appropriate limiting procedure).
In order to identify this surface, let us analyze its intrinsic
geometry. Using the defining relation (\ref{stau})
it is not difficult to obtain the first fundamental form in $S^{FL}_{\tau}$,
which reads
\[
ds^2_{S^{FL}_{\tau}} = \left .
\frac{\Sigma^2}{1- \sigma_0^2 \sin^2 \theta} d\theta^2 + \Sigma^2
\sin^2 \theta d\phi^2 \right |_{r=r_{\tau} \left (\theta \right)}.
\]
Evaluating the curvature scalar $R_{S^{FL}_{\tau}}$ for this metric we find
\begin{equation}
R_{S^{FL}_{\tau}} = \frac{
\left [\left (1+\sigma_0 \right)^2 + \epsilon \sigma_1^2 \right]
\left [\left (1- \sigma_0 \right)^2 + \epsilon \sigma_1^2 \right ]}{2
\sigma_1^2},
\label{curvature}
\end{equation}
which is constant throughout $S^{FL}_{\tau}$. Thus, this two-surface
is maximally symmetric and hence \cite{Eis}
it is either a  two-sphere, a two-plane or a hyperbolic plane
depending on whether the sign of $R_{S^{FL}_{\tau}}$
is positive, zero or negative, respectively. Due to (\ref{curvature}), it
follows that in a closed FL spacetime ($\epsilon=1$) the matching
two-surface $S^{FL}_{\tau}$ must be a two-sphere,  in a
flat FL cosmology ($\epsilon=0$)
both two-spheres and two-planes are possible
and in an open FL cosmology ($\epsilon=
-1$) the three possibilities above are, in principle, allowed.
The physical problem we are investigating is the existence of axially symmetric
regions immersed in a Friedman-Lema\^{\i}tre background.  When the
surfaces $S^{FL}_{\tau}$ are two-planes or hyperbolic planes, the
static regions are
spatially unbounded (they extend to the spatial infinity in the
FL cosmology)
and the only possible construction for having a static
region in a FL background would be performing a
double matching, so that the static region is ``sandwiched'' between two
FL spacetimes (i.e. the static region would have finite width but would be
unbounded in the other two spatial directions). Even in that
case, the static region cannot be prevented to disconnect
the FL background. Thus, these two cases are not
physically interesting for the problem we are considering in this paper.
Consequently, we will restrict the analysis to the case in
which $\{ S^{FL}_{\tau} \}$ are two-spheres (i.e (\ref{curvature})
is always positive).

Let us now locate these two-spheres
within the FL spacetime. To that end, we take advantage of the
high degree of symmetry in the cosmological background. The
uniparametric group of transformations generated by the Killing
vector
\begin{equation}
\vec{k}= \cos \theta \partial_r - \frac{\Sigma_{,r}}{\Sigma} \sin \theta
\partial_{\theta} \label{kilvec}
\end{equation}
is the most general translational isometry leaving the symmetry
axis of $\partial_{\phi}$ invariant.
Let us use the natural parametrization (which will be denoted by $\gamma$)
induced by (\ref{kilvec})
in this group of transformations. It is not difficult to show
that the transformation $\gamma=\gamma_0$, where
$\gamma_0$ is given by
\[
\Sigma(\gamma_0,\epsilon) = \frac{2\sigma_1 \sigma_0}{\sqrt{ \left [
\left (1+\sigma_0\right)^2
+\epsilon \sigma_1^2 \right ] \left [  \left (1-\sigma_0\right)^2+
\epsilon \sigma_1^2
\right ]}},
\]
transforms the surface (\ref{stau}) into the two-sphere $r= r_0$, with
$r_0$ given by
\[
\Sigma \left (r_0,\epsilon \right) =
\frac{2\sigma_1}{\sqrt{ \left [ \left (1+\sigma_0\right)^2
+\epsilon \sigma_1^2 \right ] \left [   \left (1-\sigma_0\right)^2+
\epsilon \sigma_1^2
\right ] }}.
\]
Thus, the matching hypersurface in the FL cosmology can be,
at most, a two-sphere moving arbitrarily along the axis of symmetry
and with a radius which can change freely as time evolves.
At this stage, the result in
\cite{RS} stating that no cylindrically symmetric Einstein-Straus models
can exist, has been recovered and generalized.  In those models the
matching hypersurface consists of a collection of cylinders with arbitrarily
varying radius. The results above show that
any model in which the matching hypersurface at
any instant of cosmic time is not a two-sphere, a two-plane or a
hyperbolic plane is {\it impossible}.

In order to complete the matching procedure, it is convenient to introduce
new coordinate systems in the FL and static spacetimes so that the
form of the matching hypersurfaces becomes simpler. To that end,
we choose  spherical coordinates on each hypersurface $t= \tau$
in the FL spacetime so that the two-sphere $S^{FL}_{\tau}$ is centred
at the origin. Completing these coordinates with
the global time $t$, we obtain a well-defined system of coordinates.
In order to obtain the line-element in
the new coordinate system, we first change the original
coordinates (\ref{FLmetric}) into cylindrical coordinates
$\{ t, \rho, z, \phi \}$ by the standard transformation
\begin{equation}
\Sigma \left (\rho, \epsilon\right) = \Sigma \left (r, \epsilon
\right) \sin \theta,
\hspace{1cm}
\frac{\Sigma \left (z, \epsilon \right)}{
\Sigma_{,z} \left (z,  \epsilon \right)} = \frac{\Sigma \left (r, \epsilon
\right)}{\Sigma_{,r} \left (r, \epsilon \right)} \cos \theta.
\label{transcilesf}
\end{equation}
($t$ and $\phi$ remain unchanged). In cylindrical coordinates,
the Killing vector (\ref{kilvec})  reads
simply $\vec{k} = \partial_z$ and the corresponding
uniparametric group of transformations takes the trivial
form
\begin{equation}
t \rightarrow t, \hspace{1cm} \rho \rightarrow\rho, \hspace{1cm}
 z \rightarrow z + \gamma_0(t), \hspace{1cm} \phi \rightarrow \phi,
\label{tracyl}
\end{equation}
where $\gamma_0$ is allowed to change in time because
the centre of the two-sphere $S^{FL}_{\tau}$ can move along the
symmetry axis. Applying (\ref{tracyl})
and then transforming back
into spherical coordinates by using (\ref{transcilesf}), we obtain
the line-element
\begin{eqnarray}
ds^{2}_{FL} = -dt^2 + a^2 \left (t \right) \left [
\left ( \frac{}{}
dr + f(t) \cth dt \right)^2  + \left (
\frac{}{} \Sigma  \left (r, \epsilon \right)
d \theta  - f(t)  \Sigma_{,r} \left (r,\epsilon \right) \sin \theta
 dt \right )^2
\right . \nonumber \\
\left . \hspace{7cm} + \Sigma^2  \left (r, \epsilon \right) \sin^2 \theta
d {\phi}^2  \frac{}{}\right]. \label{FLnew}
\end{eqnarray}
where $f(t) \equiv \gamma_0(t)_{,t}$ denotes the motion of the centre
of the two-sphere along the symmetry axis and, for notational simplicity,
the same symbols
$\{t,r,\theta,\phi \}$ are used to name
the new coordinates (which will be used from now on).
The embedding defining  the matching surface takes now the very
simple form
\begin{eqnarray}
\Phi^{FL} : \, \, \, \Omega \, & \longrightarrow & \, \, \,V^{FL}
\nonumber \\
\Phi^{FL} : \, \left \{ \tau,\vartheta, \varphi \right \}
\, & \longrightarrow & \, \left \{t= \tau, \, \,
r = r \left (\tau\right) ,
\, \,  \theta = \vartheta , \, \, \phi = \varphi \right \}, \label{hipFL}
\end{eqnarray}
and the embedding $\Phi^{FL}_{\tau} : \, S_{\tau}
 \, \rightarrow  \,  V^{FL}$ is obtained
simply by fixing $\tau= \mbox{const.}$ above.

A convenient coordinate system in the static region can be constructed as
follows. Due to the
staticity of the spacetime, the integral lines of the static Killing
$\partial_T$ define a canonical diffeomorphism between
different constant time hypersurfaces \cite{Ger}. Thus, there exists
a natural projection
\[
\Pi : \, \, \, V^{St} \, \hspace{3mm} \longrightarrow \hspace{3mm}
\, \, \, \Xi,
\]
where $\Xi$ is the quotient space of $V^{St}$ defined
by the orbits of the static Killing vector (it is, therefore,
diffeomorphic to any of the hypersurfaces $\left \{
T = \mbox{const.} \right \}$). Finding a coordinate system in an open
subset ${\cal U} \subset \Xi$ amounts to finding a coordinate system in the
four-dimensional open set $\Pi^{-1} \left ({\cal U} \right)$
because the static time $T$ can be used as the fourth coordinate.
Let us now consider the projection along $\Pi$ of the
two-spheres $\{ S^{St}_{\tau} \} $. Since $S^{St}_{\tau}$ is completely
contained in the hypersurface $\{ T = T(\tau) \}$
it follows that $S^{St}_{\tau}$ is isometric to
$\Pi \left (S^{St}_{\tau} \right)$ which,
consequently, must be a two-sphere in $\Xi$. Furthermore, $\Xi$
is axially symmetric (because of the
axial symmetry of $V^{St}$)
and each $\Pi \left (S^{St}_{\tau} \right )$ is centred on the
symmetry axis of $\Xi$ (this follows from
the fact that $\{ S^{St}_{\tau}\}$ are invariant under the action of the
axial symmetry in $V^{St}$). Defining the subset of $\Xi$
\[
{\cal W} = \bigcup_{\tau \in J} \Pi \left ( S^{St}_{\tau} \right),
\]
where $J$ is the interval of variation of the parameter $\tau$, it is
clear that either ${\cal W}$ is also a two-sphere or ${\cal W}$ has a
non-empty interior. The first situation is marginal and it only happens
when all the two-spheres $\{ S^{St}_{\tau} \}$ are projected
to each other by the
integral curves of $\partial_{T}$.  Consequently, they must all
have the same radius, which translates into
$\Sigma( r(\tau) ) a(\tau) = \mbox{const.}$ in the FL background. This
fixes $r(\tau)$ once the scale factor is given.
Furthermore, a trivial calculation shows
that the matching conditions (\ref{Mat1}) imply $f \equiv 0$.
Thus, the matching hypersurface in the FL spacetime
is completely fixed and must be a two-sphere of constant geometrical
radius with its centre being at rest with respect to the cosmological
flow. Hence, the shape of the static cavity {\it must} be spherically
symmetric in this case.

In the generic situation, the set of matching
two-spheres $\{ S^{St}_{\tau} \}$ are not all projected to each
other by the integral curves of $\partial_T$ and, hence,
${\cal W}$ has a non-empty interior $\dw$.
At any point $p \in \dw$ there exists an open
neighbourhood $U_p \subset \dw$ and a sufficiently
small open subinterval $I_p \subset J$ such that
the set of two-spheres $\left \{ \Pi \left (S^{St}_{\tau}\right ),
 \, \,  \tau \in I_p \right \}$ foliates $U_p$. In other words,
for any point $q \in U_{p}$ there
exists a single value of $\tau_{q} \in I_{p}$ such that $q$ belongs to
$\Pi \left (S^{St}_{\tau_q} \right)$. A coordinate system
in $U_p$ can now be constructed as follows. First, choose the natural angular
coordinates $\{ \theta, \hat{\phi} \}$ on each
two-sphere $\Pi \left (S^{St}_{\tau} \right)$ such that
$\theta=0$ and $\theta = \pi$ correspond to the axial symmetry axis of
$\Xi$ (this choice is possible because
each two-sphere $S^{St}_{\tau}$ is centred with respect to the axis of
symmetry in $V^{St}$). The third coordinate necessary to
identify any point $q \in U_{p} \subset \Xi$ can be chosen as the unique
value $\tau_q \in I_p$ identifying the two-sphere passing through $q$. In
this way, a coordinate system around any point $p \in \dw$ is constructed.
This coordinate system cannot, in general, cover the whole set $\dw$
(because two different two-spheres with arbitrarily close values of
$\tau$ may intersect). Due to the purely local character of the
matching conditions, this is not an obstacle to proceed with the
matching. In order to simplify the matching equations and the final
form of the interior static metric obtained below, it turns out to be
convenient to use the symbol $t$ to denote the coordinate identifying the
two-sphere passing through any point $q \in U_{p}$. This coordinate
is spacelike by construction and it has nothing to do with the cosmic time
in the FL background so that they should not be confused. The meaning of $t$
should become clear from the context. The reason why the same symbol is used
for such different objects is that both take the same value at every point on
the matching hypersurface, which allows for an important simplification in
the writing of the expressions below.

The line-element for the  static spacetime in the coordinate
system just constructed takes the form
\begin{equation}
ds^2_{St} = - F^2 \left (t, \theta \right) dT^2 + Q^2 \left (t, \theta \right)
dt^2 + \hat{r}^2 \left (t \right) \left [ \left (\frac{}{}
d \theta + H \left( t, \theta
\right ) \, dt \right )^2 +  \sin ^2 \theta d\hat{\phi}^2
\right ], \label{metSt}
\end{equation}
which is the most general metric for a static, axially
symmetric and orthogonally transitive spacetime such that the
two-surfaces $T=\mbox{const}$, $t= \mbox{const.}$ are two-spheres.

The embedding of
the matching hypersurface $\Omega$ into $V^{St}$ takes now
the simple form
\begin{eqnarray}
\Phi^{St} : \, \, \, \Omega \, & \longrightarrow & \, \, \,V^{St}
\nonumber \\
\Phi^{St} : \, \left \{ \tau, \vartheta, \varphi  \right \}
\, & \longrightarrow & \, \left \{T = T(\tau), \, \,
t = \tau ,
\, \,  \theta = \vartheta , \, \, \hat{\phi} = \varphi \right \}
\label{hipSt}
\end{eqnarray}
and the corresponding embeddings
$\Phi^{St}_{\tau} : \, S_{\tau} \,  \rightarrow  \,
V^{St}$ are obtained by fixing $\tau = \mbox{const.}$

We are now in a favourable position to complete the matching problem.
We already made use
of the constraint conditions (\ref{ConstMat1}) and (\ref{ConstMat2})
in order to restrict the form of $\Omega$.
However, not all their information
has been extracted yet.
The equality of the two first fundamental
forms on $S_{\tau}$ gives now the single relation
\begin{equation}
 \hat{r} \left(t  \right) =
\left . a (t) \Sigma
\left (r, \epsilon \right) \frac{}{}\right |_{r = r (t )}. \label{FConF}
\end{equation}
The vector spaces $\left (T_{x} S_{\tau}^{FL} \right )_{\bot}$ and
$\left (T_{x} S_{\tau}^{St} \right )_{\bot}$ are
\[
\left (T_{x} S_{\tau}^{FL} \right )_{\bot} = \left <
\left . dt \right |_{\F}, \left . dr \right |_{\F} \right >,
\hspace{1cm}
\left (T_{x} S_{\tau}^{St} \right )_{\bot} = \left <
\left . dT \right |_{\St}, \left . dt \right |_{\St} \right >
\]
and the most general linear isometry $\xi^{x}_{\tau}$ between them
is the linear extension of
\begin{eqnarray}
\xi^{x}_{\tau}\left (\left . dt \right |_{\F}
\right) = \left . \eta_1 \cosh \beta F dT +
\eta_2 \sinh \beta Q dt \frac{}{}
\right |_{\St}, \nonumber\\
\xi^{x}_{\tau}\left ( \left . dr\right |_{\F}
 \right) =   \eta_1 F \left ( \frac{\sinh \beta}{a}
- f \cos \theta \cosh \beta \right ) dT + \hspace{4cm} \label{xit} \\
\hspace{6cm} + \left . \eta_2 Q \left ( \frac{\cosh
\beta}{a} -   \frac{}{} f \cos \theta \sinh \beta \right ) dt
\frac{}{} \right |_{\St}.
\nonumber
\end{eqnarray}
A straightforward calculation shows that (\ref{ConstMat2})
are equivalent to the three equations
\begin{eqnarray*}
\left . \sin \theta H_{,\theta} - \cos \theta H \right |_{S_{\tau}}=0 , \\
\left . \eta_2 \hat{r} \sinh \beta  \left (   H \hat{r} \cos \theta -
\hat{r}_{,t} \sin \theta \right ) - Q \Sigma^2 a a_{,t}  \sin \theta
\right |_{S_{\tau}}=0 ,  \\
\left . \Sigma a_{,t}  \cosh \beta +  \Sigma_{,r} \sinh \beta
 \right |_{S_{\tau}}=0 ,
 \end{eqnarray*}
which can be trivially solved to give
\begin{eqnarray}
\left . H \left ( t , \theta \right)  = L \left (t \right) \sin \theta,
\hspace{15mm}  Q\left (t,\theta \right)
= \frac{\hat{r}_{,t} -
L \hat{r} \cos \theta }{\eta_2 \sqrt{\Sigma^2_{,r}
 - \Sigma^2 a_{,t}^2}} \right |_{r=r(t)} , \label{HQ}\\
\sinh \beta = \left . \frac{-\Sigma a_{,t}}{ \sqrt{\Sigma^2_{,r}
 - \Sigma^2 a_{,t}^2}}
 \right |_{r=r(t)}, \hspace{15mm}
\cosh \beta = \left . \frac{ \Sigma_{,r}}{\sqrt{\Sigma^2_{,r}
 - \Sigma^2 a_{,t}^2}} \right |_{r=r(t)}, \label{sinhcosh}
\end{eqnarray}
where $L(t)$ is an arbitrary function. Thus, the
$\theta$ dependence of the
two metric coefficients $Q$ and $H$ in (\ref{metSt})
is completely fixed. These expressions show that a necessary condition
for the existence of the matching is
\begin{equation}
\left . \Sigma^2_{,r}  - \Sigma^2 a_{,t}^2\frac{}{} \right |_{r=r(t)} > 0.
\label{NeCon}
\end{equation}
This inequality can be geometrically interpreted by noticing
that the two-sphere $S_{t}^{FL}$ is non-trapped (see e.g. \cite{HE}
for a definition) if and only if (\ref{NeCon}) is fulfilled.
Thus, it follows that a trapped (or
marginally trapped) surface cannot be the boundary of an axially
symmetric static region, which is an intuitively clear result.
Using the expression (\ref{enerFL}) below
for the cosmic energy-density $\rho^{FL}$,
the inequality (\ref{NeCon}) can be rewritten as
\begin{equation}
\Sigma \left (r(t),\epsilon \right) <
\frac{1}{a(t)} \sqrt{\frac{3}{\rho^{FL}(t)}}. \label{NeCon2}
\end{equation}
Therefore, the radius of the static region must be smaller than some
maximum radius. In particular, this forbids situations in which the
radius of the two-sphere $S^{FL}_{\tau}$
grows unboundedly until reaching infinity
(so that, in fact, it becomes a two-plane).
Hence, if the matching hypersurface is a two-sphere at any instant of
time, it must remain a two-sphere for
ever (of course, this two-sphere can become a point and then
disappear).

Having exhausted the constraint matching conditions, we must now
consider the more complicated set of conditions (\ref{Mat1}) and
(\ref{Mat2}). The task is, however, much simpler now because
of all the information we already know. The equality of the two first
fundamental forms on $\Omega$ provides
\begin{equation}
 L  = - \left . \frac{\Sigma_{,r}}{\Sigma} \right |_{r=r(t)} f(t),
\hspace{15mm}
F T_{,t}    =
\left . \frac{\Sigma_{,r} + a  a_{,t} \Sigma \left (r_{,t} + f \cos \theta
\right )}{\sqrt{\Sigma^2_{,r}
 -  a_{,t}^2 \Sigma^2}} \right |_{r=r(t)}. \label{LF}
\end{equation}
which fixes the $\theta$ dependence of the remaining metric
coefficient $F$ in the static spacetime.
Using (\ref{FConF}), (\ref{HQ}) and (\ref{LF}),
the static line-element can be rewritten in the form
\begin{eqnarray}
ds^2_{St} = - \frac{\left [ \Sigma_{,r} +  a a_{,t} \Sigma \left ( r_{,t}
+ f \cos \theta \right ) \right ]^2}{T_{,t}^2 \left(
\Sigma_{,r}^2 - a_{,t}^2 \Sigma^2\right) } dT^2
+\frac{\left [ a_{,t} \Sigma + a \Sigma_{,r} \left ( r_{,t}
+ f \cos \theta \right ) \right ]^2}{
\Sigma_{,r}^2 -  a_{,t}^2 \Sigma^2  } dt^2 + \nonumber \\
\left . + a^2 \Sigma^2 \left (d \theta
-\frac{\Sigma_{,r}}{\Sigma} f \sin\theta dt \right)^2 + a^2 \Sigma^2
\sin^2 \theta d \hat{\phi}^2 \right |_{r=r \left (t \right)},\label{metst}
\end{eqnarray}
which is completely given in terms of cosmological quantities, except
for the function $T(t)$ (which fixes the form of the matching hypersurface
as seen from the static region). In order to impose the second set of matching
conditions we choose the rigging
\begin{eqnarray*}
\vec{l}_{FL} = \frac{1}{a} \partial_r \hspace{5mm} \Longrightarrow
\hspace{5cm}\\
\left . \bm{l_{FL}} \frac{}{} \right |_{\Omega^{FL}}
 = \left . a \left (f \cos \theta dt + dr \frac{}{} \right)
 \right |_{\Omega^{FL}},
\hspace{4mm}
\left . \bm{l_{St}} \frac{}{} \right |_{\Omega^{St}}
 =\left . \frac{}{} \eta_1 F \sinh \beta dT + \eta_2 Q \cosh \beta dt
\right |_{\Omega^{St}},
\end{eqnarray*}
where the one-form $\bm{l_{St}}$ has been obtained from
$\bm{l_{FL}}$ by applying the linear isometry (\ref{xit}).
Evaluating (\ref{Mat2}) involves now a
rather long calculation which was performed with the aid of the
computer algebra program REDUCE \cite{MW}. Only the $(\tau, \tau)$ component
is not identically satisfied and it provides two conditions.
Whenever $f$ is not identically vanishing these two equations
can be rewritten as
\begin{eqnarray}
T_{,tt} f = T_{,t}\left [ f_{,t} + 2 f
\left . \frac{a_{,tt} \Sigma_{,r}^2 + a_{,t} \Sigma \Sigma_{,r} r_{,t}
\left (\epsilon + a_{,t}^2 \right )}{
a_{,t} \left (\Sigma_{,r}^2 - a_{,t}^2 \Sigma^2 \right)} \right |_{r=r(t)}
\right ], \label{eqdotT}\\
\left ( a a_{,t} \Sigma r_{,t} + \Sigma_{,r} \right ) a_{,t} f_{,t}
 + a a_{,t}^2 \Sigma_{,r} f^3+ \hspace{65mm} \nonumber \\
\hspace{2cm}
\left . +  \left [ -a a_{,t}^2 \left ( \Sigma r_{,tt} +
\Sigma_{,r} r_{,t}^2 \right )
+ 2 \left ( a_{,tt} \Sigma_{,r} + \epsilon  a_{,t} \Sigma r_{,t} \right)
\right ]f  \right |_{r=r(t)} = 0.
 \label{eqf}
\end{eqnarray}
The case $f\equiv 0$ can be obtained from these equations
by substituting $f_{,t}$ from (\ref{eqf}) into (\ref{eqdotT}), then dropping
a global factor $f$ and finally putting $f=0$. Only one equation
survives in that particular case.

The expressions (\ref{eqdotT}) and (\ref{eqf})
are differential equations for $T(t)$ and $f(t)$. Thus, they
determine the form of the matching hypersurface in the static spacetime
and the motion of the centre of the two spheres $\{S^{FL}_{t} \}$ along the
axis of symmetry in the FL spacetime.
Both equations can be explicitly integrated once the scale factor
$a(t)$  and the evolution $r(t)$
of the radius of the two-spheres $\{ S^{FL}_{t} \}$ are given.
In order to solve (\ref{eqf}) two different cases must be considered.
In the particular case
\begin{equation}
\left .  \Sigma_{,r} + a a_{,t} \Sigma r_{,t} \right |_{r=r(t)} \equiv 0
\label{partcase},
\end{equation}
the equation becomes algebraic and its solution is
(the other solution $f\equiv 0$ is impossible due to (\ref{LF}))
\[
f^2(t) = \frac{a_{,t}^2 + \epsilon - a a_{,tt}}{a^2 a_{,t}^2} =
\frac{\left (\rho + p \right)^{FL}}{2 a_{,t}^2},
\]
where the last equality follows from the standard expressions for the
energy-density and pressure in the FL spacetime in terms of the scale factor
(see (\ref{enerFL}) below).

In the generic case, (\ref{partcase}) does not hold and (\ref{eqf}) is
a Bernouilli differential equation
which can integrated \cite{Kam} in terms of quadratures as
\begin{equation}
f(t) = \frac{ \alpha e^{\int{f_1 dt }}}{
\sqrt{1 - 2 \alpha^2 \int{ \frac{}{} f_2 e^{2 \int{ f_1 dt}} dt}}},
\label{solf}
\end{equation}
where $\alpha$ is an arbitrary constant  and $f_1(t)$, $f_2(t)$
are defined as
\[
f_1 (t) \equiv \left .\frac{a a_{,t}^2 \left ( \Sigma r_{,tt} + \Sigma_{,r}
r_{,t}^2 \right )
- 2 \left ( a_{,tt} \Sigma_{,r} + \epsilon  a_{,t} \Sigma r_{,t} \right)}{
a_{,t} \left ( a a_{,t} \Sigma r_{,t} + \Sigma_{,r} \right ) }
\right |_{r=r(t)},
f_2(t)  \equiv \left . \frac{ - a a_{,t} \Sigma_{,r}}{a a_{,t} \Sigma r_{,t} +
\Sigma_{,r}} \right |_{r=r(t)}.
\]
Regarding (\ref{eqdotT}), this is a linear differential equation for
$T_{,t}$  which is therefore trivial to  integrate. The structure
of its general solution is $T(t) = \alpha_1 T_0 (t) + \alpha_0$,
where $T_0(t)$ is any particular solution and $\alpha_0$,  $\alpha_1$ are
integration constants.
These constants are irrelevant as
the form of the matching hypersurface remains unaffected if they change.
They simply correspond to the freedom (\ref{freedSt})
in defining the global static time in any static spacetime. Thus,
$\alpha_0$ and $\alpha_1$ can be fixed arbitrarily without loss of generality.

Thus, we  conclude that the matching of a Friedman-Lema\^{\i}tre
spacetime with an axially symmetric static region allows for an
arbitrary scale factor in the FL spacetime and for an arbitrary
function $r(t)$ describing the evolution of the radius of the matching
two-spheres (as long as the inequality (\ref{NeCon2}) is fulfilled).
Then, the matching conditions fix the motion of the centre
of the two-spheres along the axis of symmetry through (\ref{solf}). They
also fix uniquely the form of the matching hypersurface in
the interior static region {\it and } the full geometry of the
static spacetime (in an open region around the matching hypersurface),
which is given by the line-element (\ref{metst}).
In the next section, the geometry of the interior static geometry will be
analyzed.

\section{Geometry of the static spacetime.}

Let us first of all discuss the domain of applicability of the
coordinate system  $\{ T,t, \theta, \hat{\phi} \}$
in the static spacetime. As remarked in
the previous section, these coordinates are well-defined at
least in a sufficiently small open neighbourhood around any point
in the interior of $\Pi^{-1} (\Pi  (\Omega^{St}))$.
A priori, there is no reason to expect this
coordinate system to cover the whole open set
$\Pi^{-1} (\dw )$. However, this turns out to be the case
and a single coordinate patch can be used to describe
all the relevant interior
region. Before showing this explicitly, it is worth pointing out
that the form of the interior
static metric at points {\it not} belonging to $\Pi^{-1}(\dw)$ can not
be determined at all by the matching conditions. This is because the
matching procedure is purely local and there is no way of relating points
outside this open set with the matching hypersurface. Determining the
interior metric at those points requires further information like
analyticity conditions or the Einstein field equations for some
type of energy-momentum contents. Thus, all the discussion below concerning
the geometry of the static region is restricted to $\Pi^{-1}(\dw)$.

In order to see explicitly that
$\{ T,t, \theta, \hat{\phi} \}$ is a valid coordinate system
everywhere,
we must show that no coordinate singularities are present in the
line-element (\ref{metst}) or, equivalently, we must proof that any
possible coordinate singularity is, in fact, an essential singularity.
The metric (\ref{metst}) becomes singular at any point satisfying
\begin{equation}
\Sigma_{,r} +  a a_{,t} \Sigma \left ( r_{,t}
+ f \cos \theta \right ) = 0 \hspace{1cm} \mbox{or} \hspace{1cm}
 a_{,t} \Sigma + a \Sigma_{,r} \left ( r_{,t}
+ f \cos \theta \right )=0. \label{singu}
\end{equation}
The remaining possibilities, like $T_{,t}=0$ or $f \rightarrow +\infty$,
have clear implications for the matching hypersurface
as seen from the FL spacetime (where the coordinates have
a transparent meaning) and can be discarded as unphysical for the problem
we are considering.
In order to show that points satisfying (\ref{singu}) are
essential singularities, let us write down the energy-momentum tensor
of the static spacetime. Using the orthonormal tetrad
\begin{eqnarray}
\bm{\theta^0} = \left . \frac{ \Sigma_{,r} +  a a_{,t} \Sigma \left ( r_{,t}
+ f \cos \theta \right )}{T_{,t}  \sqrt{
\Sigma_{,r}^2 - a_{,t}^2 \Sigma^2 } } dT \right |_{r=r \left (t \right)},
 \hspace{3mm}
\bm{\theta^1} = \left . \frac{a_{,t} \Sigma + a \Sigma_{,r} \left ( r_{,t}
+ f \cos \theta \right ) }{
\sqrt{\Sigma_{,r}^2 - a_{,t}^2 \Sigma^2} } dt \right |_{r=r\left( t \right)},
\label{tetrad} \\
\bm{\theta^2} = \left . a \Sigma  \left (d \theta
-\frac{\Sigma_{,r}}{\Sigma} f \sin \theta dt \right) \right |_{r=r
\left ( t \right )}, \hspace{3mm}
\bm{\theta^3} =  \left . a \Sigma
\sin \theta d \hat{\phi} \right |_{r=r\left( t \right)}  \nonumber
\end{eqnarray}
to evaluate all tensor components, it turns out that the energy-momentum
tensor is  diagonal and with the structure
$( T_{\alpha \beta}) = \mbox{diag} \left ( \rho, p_r, p_T, p_T \right)$,
where $\rho$ stands for the energy-density, $p_r$ is the radial
pressure and $p_T$ the tangential pressure. We borrow the
terms radial and tangential pressure from the spherically
symmetric case even though the static metric does not possess this
exact symmetry
in general. However, the static metric is nearly
spherically symmetric (its departure being
 governed by the single function $f$) since at every point we have
a two-sphere passing through it.
The structure of the energy-momentum tensor is exactly the
same as in the exact spherically symmetric case.  This resemblance
goes even further because the evaluation of the Weyl spinor in the null tetrad
associated canonically with (\ref{tetrad}) shows that $\Psi_2$
is the  only non-vanishing component, exactly as in the spherically
symmetric case. Thus, the static spacetime is algebraically special of
Petrov type D (or conformally flat when $\Psi_2$=0).

The explicit expressions for $\rho$ and $p_r$ can be
read from the expressions
\begin{eqnarray}
\rho = \left .
\frac{ 3 a \Sigma_{,r} \left (a_{,t}^2 + \epsilon \right) \left (f \cos \theta
+ r_{,t} \right) + a_{,t} \Sigma \left (2 a a_{,tt}
+ a_{,t}^2 + \epsilon \right)}{ a^2 \left [
a_{,t} \Sigma + a \Sigma_{,r} \left ( r_{,t}+ f \cos \theta \right )
\right ]} \right |_{r=r \left (t \right)}, \label{rhoest}\\
\rho + p_r = \left .
\frac{ a \left ( \rho + p \right)^{FL}
\left ( \Sigma_{,r}^2 - a_{,t}^2 \Sigma^2 \right)
\left ( f \cos \theta + r_{,t} \right) }{
 \left [ a_{,t} \Sigma + a \Sigma_{,r} \left ( r_{,t}+ f \cos \theta \right )
\right ] \left [  \Sigma_{,r} +  a a_{,t} \Sigma \left ( r_{,t}
+ f \cos \theta \right )
\right ] } \right |_{r = r \left ( t \right)}, \label{rhopreest}
\end{eqnarray}
where $\rho^{FL}$ and $p^{FL}$ stand for the cosmic
energy-density and pressure in terms of the scale factor
\begin{equation}
\rho^{FL} = \frac{3 \left (a_{,t}^2 + \epsilon \right)}{a^2},
 \hspace{15mm} p^{FL} =  \frac{-2 a a_{,tt} - a_{,t}^2 - \epsilon}{a^2}.
\label{enerFL}
\end{equation}
Regarding $p_T$, its expression takes different forms depending on
whether $a a_{,t} \Sigma r_{,t} + \Sigma_{,r} \neq 0$ or not. In the first
case we obtain
\begin{eqnarray}
\rho+p_T =
\frac{\left (\rho+p \right)^{FL}_{,t} a \Sigma \Sigma_{,r}
+ 2 \left (\rho+ p \right)^{FL} a \Sigma_{,r}^2 \left (r_{,t} + f \cos \theta
\right)}{2  \left [ a_{,t} \Sigma + a \Sigma_{,r} \left ( r_{,t}+ f \cos \theta \right )
\right ] \left [  \Sigma_{,r} +  a a_{,t} \Sigma \left ( r_{,t}
+ f \cos \theta \right )
\right ]} +
\left ( \rho + p \right)^{FL} \Sigma \times \nonumber \\
\left . \times \frac{a \Sigma \Sigma_{,r} r_{,t}
 \left (
3 a_{,t}^2 - 2 a a_{,tt} \right) + a_{,t} \left [ \Sigma_{,r}^2 \left (
3 + a^2 f^2 \right) - a^2 \left (
1+\epsilon \Sigma^2 \right) r_{,t}^2 - a^2 \Sigma \Sigma_{,r} r_{,tt}
 \right ]}{
2  \left [ a_{,t} \Sigma + a \Sigma_{,r} \left ( r_{,t}+ f \cos
\theta
 \right )
\right ] \left [  \Sigma_{,r} +  a a_{,t} \Sigma \left ( r_{,t}
+ f \cos \theta \right )
\right ]
\left [ \Sigma_{,r} + a a_{,t} \Sigma r_{,t} \right ] }
\right |_{r=r \left ( t \right)} \label{rhopreTest}
\end{eqnarray}
and in the second case ($a a_{,t} \Sigma r_{,t} + \Sigma_{,r} \equiv 0$),
the corresponding expression reads
\begin{eqnarray*}
\rho+p_{T}=  \frac{1}{4a a_{,t}
\Sigma f \cos \theta \left (
a_{,t}^2 \Sigma^2 - \Sigma_{,r}^2 + a a_{,t} \Sigma \Sigma_{,r} f \cos \theta
\right)} \left ( \frac{}{}
a a_{,t} \Sigma^2 \Sigma_{,r} \left (\rho
+ p \right)^{FL}_{,t} + \right . \hspace{5mm} \\
+ \left . \left . 2 \left (\rho+p \right)^{FL} \Sigma_{,r}
\left [2a a_{,t} \Sigma \Sigma_{,r} f \cos \theta + \Sigma^2
\left (3 a_{,t}^2 + 4 \epsilon - a a_{,tt} \right) - 2 \right ]
\frac{}{}\right ) \right |_{r=r(t)}.
\end{eqnarray*}

Since the scalar quantities $\rho$, $p_r$ and $p_T$ have a clear physical
meaning, any divergence in them corresponds to a true singularity
in the static metric. It is now clear that at any point satisfying
(\ref{singu}) we have divergences in $\rho$ and/or $p_r$ and, therefore,
we have a matter singularity rather than just
a bad coordinate description. This type of singularities leads to
physically unrealistic models and should, therefore, be discarded
(notice that these singularities are ``touched'' by the matching hypersurface
and cannot be interpreted as singular ``sources'' for the static metric,
like in the Schwarzschild spacetime)

The expressions for $\rho$, $p_r$ and $p_T$ above
allow also for obtaining a number of interesting conclusions.
When the FL spacetime is de Sitter (or anti-de Sitter)
then $(\rho+p)^{FL} = 0$ and the energy-momentum tensor in the
static spacetime becomes $T_{\alpha \beta} = \Lambda g^{St}_{\alpha \beta}$
(with the same value for $\Lambda$ as
the cosmic background). Furthermore, the static spacetime can be easily
seen to be conformally flat and therefore it must necessarily be
de Sitter (anti-de Sitter) \cite{KSMH}. Thus, the following result holds
\begin{proposition}
Any static and axially symmetric
spatially compact region in a de Sitter background must have
the same geometry as the exterior manifold.
\end{proposition}
In other words, de Sitter (anti-de
Sitter) spacetimes do not admit non-trivial static and axially symmetric
spatially compact regions. Thus, we can assume from now on that
that the FL spacetime is not de Sitter or anti-de Sitter, i.e. $\left (
\rho+p \right )^{FL} \neq 0$.

If the static region is imposed to have an energy-momentum tensor
with $\rho+p_r =0$ (this includes the important subcases of
vacuum, $\Lambda$-term and
non-null Einstein-Maxwell solutions, see below),
then (\ref{rhopreest}) immediately implies $f=0$ and
$ r_{,t}= 0$ ($\Longleftrightarrow r(t) = r_0$).
Thus, the matching hypersurface must be a
two-sphere comoving with the cosmological flow. Conversely, if
the matching two-sphere is comoving with the cosmological flow
($f=0$ , $r(t)=r_0$), then (\ref{rhopreest}) implies $\rho+p_r =0$, which
shows the equivalence of the two conditions.
In this situation, the differential
equation for $T_{,tt}$ (\ref{eqdotT}) can be integrated to give
\[
T_{,t}= \left .
\frac{\Sigma_{,r}}{\Sigma_{,r}^2 - a_{,t}^2 \Sigma^2} \right |_{r=r_0}
\]
(the integration constants are irrelevant, as discussed above) so that
the line-element in the static region becomes
\begin{equation}
ds^2 = \left . - \left ( \Sigma_{,r}^2 - a_{,t}^2 \Sigma^2 \right) dT^2 +
\frac{a_{,t}^2 \Sigma^2}{\Sigma_{,r}^2 - a_{,t}^2 \Sigma^2 } dt^2 +
a^2 \Sigma^2 \left (d \theta^2 + \sin^2 \theta d \hat{\phi}^2 \right)
\right |_{r=r_0},
\label{spheest}
\end{equation}
which is obviously spherically symmetric. Performing the coordinate
change
\[
a(t) \Sigma(r_0) = \check{r}
\]
and defining the standard mass function $m(\check{r})$ \cite{Zan}
as
\[
a_{,t}^2 = - \epsilon + \frac{2 m \left ( a \Sigma \right )}{a \Sigma^3},
\]
the line element (\ref{spheest}) transforms into
\begin{equation}
ds^2 = - \left (1 - \frac{2m(\check{r})}{\check{r}} \right ) dT^2 +
\frac{d \check{r}^2}{1 - \frac{2m(\check{r})}{\check{r}}} +
\check{r}^2 \left (d \theta^2 + \sin^2 \theta d \hat{\phi}^2 \right).
\label{esphest}
\end{equation}
It is interesting to notice that this line-element is the most
general static and spherically symmetric metric with $\rho+p_r=0$.
Thus, {\it all} these metrics can be matched with an appropriate
exterior FL spacetime through a spherically symmetric matching
hypersurface. The  energy-momentum contents and the
only non-zero Weyl spinor coefficient, $\Psi_2$, of this spacetime
read
\begin{equation}
\rho = -p_r = \frac{2m_{,\check{r}}}{\check{r}^2}, \hspace{1cm}
p_T = - \frac{m_{,\check{r}\check{r}}}{\check{r}}, \hspace{1cm} \Psi_2 =
- \frac{m_{,\check{r}\check{r}}}{6 \check{r}} +
\frac{2 m_{,\check{r}}}{3 \check{r}^2} -
\frac{m}{\check{r}^3}, \label{enersph}
\end{equation}
while the energy-density and pressure in the FL
cosmology are
\[
\rho^{FL} = \frac{6m}{\check{r}^3}, \hspace{2cm} p^{FL} =
-\frac{2m_{,\check{r}}}{\check{r}^2}.
\]
Thus, we can conclude

\begin{proposition}
The matching between a static and axially symmetric spacetime
and a Friedman-Lema\^{\i}tre cosmology across a hypersurface described
by a two-sphere comoving with the cosmological flow  must
have a spherically symmetric interior with metric given by
(\ref{esphest}) where $m(\check{r})$ is an arbitrary function describing
the total mass contained in the static region.
\end{proposition}

Since vacuum and $\Lambda$-term spacetimes are particular cases
of $\rho+p_{r} =0$, the two following results hold
\begin{corollary}
The only $\Lambda$-term static and axially symmetric spatially
compact region within
a FL spacetime must be a two-sphere comoving with
the cosmological flow and interior geometry given by (\ref{esphest})
with $m= m_0 + \frac{\rho_0 \check{r}^3}{6}$ ($m_0$ and
$\rho_0$ constants).
\end{corollary}
\begin{corollary}
The only static and axially symmetric, spatially compact,
vacuum region in a
Friedman-Lema\^{\i}tre spacetime must be a two-sphere comoving with
the cosmological flow and with Schwarzschild interior geometry (the
Einstein-Straus model)
\end{corollary}
Let us next discuss
Einstein-Maxwell spacetimes, i.e. spacetimes
with an energy-momentum tensor corresponding to an
electromagnetic field outside any mass and charge sources. Static
spacetimes with a pure null electromagnetic source
are impossible \cite{Ban},
so we are left with the non-null case.
In this case (and due to the diagonal form of the energy-momentum
tensor) the so-called Rainich conditions (which are necessary
and sufficient conditions for non-null electrovacuum spacetimes)
\cite{KSMH}, imply that $(T_{\alpha\beta})$
must take the form
\[
T^{EM}_{\alpha \beta} = \mbox{diag} \left (\eta, -\eta, \eta, \eta \right).
\]
where $\eta$ is some positive quantity. Hence, $\rho+p_{r} = 0$  follows
and the metric must take the spherically symmetric
form (\ref{esphest}). The general Einstein-Maxwell solution for such
a metric is the well-known Reissner-Nordstr\"om solution,
with $m(\check{r}) = M - \frac{e^2}{2 \check{r}}$ (where $M$ is the mass and
$e$ is electric charge of the point source). The scale factor and the
energy contents in the FL spacetime are then given
by
\[
a_{,t}^2 = \left .
- \epsilon + \frac{2M}{a \Sigma^3} - \frac{e^2}{a^2 \Sigma^4}
\right |_{r=r_0},
\hspace{5mm}
\rho^{FL}  = \left . \frac{6 M}{a^3 \Sigma^3} - \frac{3e^2}{a^4 \Sigma^4}
\right |_{r=r_0},
\hspace{5mm}
p^{FL} = \left . -\frac{e^2}{a^4 \Sigma^4} \right |_{r=r_0}.
\]
Summarising, we have
\begin{proposition}
The only static and axially symmetric electrovacuum region in a
Fried\-man-Le\-ma\^{\i}tre spacetime must be a two-sphere comoving with
the cosmological flow and with Reissner-Nordstr\"om interior geometry.
\end{proposition}
Having discussed vacuum, $\Lambda$-term and electromagnetic
fields, let us next analyse the perfect-fluid case.
Imposing $p_r - p_T = 0$ and using the expressions above, we obtain
a linear relation in $\cos\theta$. The vanishing of the coefficient
in $\cos\theta$ implies necessarily $f \equiv 0$ while
the other coefficient provides a rather long equation relating
$a(t)$ and $r(t)$. Thus the following result also holds

\begin{proposition}
Any static and axially symmetric perfect-fluid region immersed
in a Friedman-Lema\^{\i}tre cosmology must be spherically symmetric
and its boundary must be an expanding (or contracting) two-sphere
with its centre being at rest with respect to the cosmological flow.
\end{proposition}
Here the terms {\it expanding} and {\it contracting} are defined with respect
to the cosmological fluid flow.

All these results above show that the most commonly used energy-momentum
tensors imply necessarily that the matching hypersurface must be a
two-sphere with its center being comoving with the cosmological fluid
(and hence the
matching hypersurface is spherically symmetric). Furthermore, the interior
geometry must also be exactly spherically symmetric
in the relevant open set $\Pi^{-1}(\dw)$ (this set may be empty
in some marginal cases - see the previous section- but
even in that particular situation the conclusions on the form of the matching
hypersurface hold). Consequently, spherical symmetry
is indeed an essential ingredient for the existence of reasonable
static cavities in a FL expanding background. This
implies that the Einstein-Straus model represents a very isolated
situation, which poses serious doubts on the
suitability of this model for providing a definitive answer to the problem
of the influence of the cosmological expansion on the local physics. Even
though this effect is probably very small at solar
system scales (there is no experimental evidence for such an effect),
this is not so clear at larger scales (ranging from galaxies
to clusters of galaxies or even larger). In any case, more adequate
models describing the problem should be analyzed before reaching definitive
conclusions on this subject.

Before finishing this paper, a final comment may be adequate. Solving any
matching problem has always a dual interpretation. Due to the symmetry $(+)
\leftrightarrow (-)$ of the matching conditions, solving a matching
between two spacetimes produces
always as a side effect the matching of the two complementary
regions discarded in the original matching (see \cite{FST} for
a clear discussion of this fact). In our case,
the complementary situation corresponds to an interior spatially
homogeneous and isotropic perfect fluid (with FL metric) and an static,
axially symmetric exterior. Thus, irrespectively of any conditions on the
static exterior (for instance, no conditions on asymptotic flatness or
similar), it turns out that the shape of the interior fluid must be a
two-sphere at each instant of time. Furthermore, for the most relevant
exterior metrics (in particular vacuum and electrovacuum), the shape of
the autogravitating fluid must be exactly spherically symmetric
and the exterior geometry must be also exactly spherically symmetric. This
side result is interesting in its own and can be phrased as follows.

A Friedman-Lema\^{\i}tre spacetime acting as a source for
static and axially symmetric spacetimes with physically
reasonable energy-momentum tensors (when the two canonical times match
properly) requires that both the boundary of the fluid and the exterior
geometry are exactly spherically symmetric, and hence the global spacetime
must also be spherically symmetric.

\section*{Acknowledgements}

I am most grateful to  Jos\'e Senovilla for a very careful reading of this
manuscript and for many useful discussions and criticisms which helped
improving this paper. I would also like to thank the Ministerio
de Educaci\'on y Cultura, Spain, for financial support under grant
EX95 40895713.

\end{document}